\documentclass[classic,doublecol]{epl2}

\usepackage[latin1]{inputenc}	
\usepackage{mathtools}
\usepackage{amssymb}
\usepackage{bm}
\usepackage{braket}

\usepackage[scaled=0.92]{helvet}
\usepackage{mathptmx}
\usepackage{microtype}

\providecommand{\mbf}{\mathbf}

\providecommand{\what}{\hat}

\RequirePackage{xspace}

\newcommand*{\viz}{\emph{viz.}\xspace}

\newcommand*{\eqn}[1]{equation~(\ref{#1})}

\renewcommand*{\d}[1]{\mathrm{d}#1}
\newcommand*{\ddt}[1]{\frac{\d{#1}}{\d{t}}}

\newcommand*{\cc}[1]{{#1}^*}		

\newcommand*{\im}{{\mathrm{i}}}		
\newcommand*{\Mag}[1]{\mathinner{\left|{#1}\right|}}

\newcommand*{\w}{\omega}
\newcommand*{\Op}[1]{\what{\boldsymbol{\mathsf{#1}}}}
\newcommand*{\op}[1]{\what{\mathsf{#1}}}

\newcommand*{\Sub}[1]{_{\text{#1}}}

\renewcommand*{\vec}[1]{\mbf{#1}}

\newcommand*{\G}{\vec{G}}
\newcommand*{\J}{\vec{J}}

\renewcommand*{\k}{\vec{k}}
\renewcommand*{\L}{\vec{L}}

\newcommand*{\p}{\vec{p}}

\renewcommand*{\S}{\vec{S}}

\title{%
Hubble and Planck scale limits on the determination of orbital\\
angular momentum states of light
}

\shorttitle{%
Hubble and Planck scale imits on the determination of OAM states of light
}

\author{%
F. Tamburini\inst{1,2}
\and
B. Thid\'{e}\inst{3,4}
\and
A. Sponselli\inst{1}
}

\shortauthor{F.\,Tamburini, B.\,Thid\'e and A.\,Sponselli}

\institute{%
\inst{1}
Department of Physics and Astronomy,
University of Padua,
Vicolo dell' Osservatorio 3,
IT-351\,22 Padua,
Italy,~EU
\\
\inst{2}
Associazione CIVEN,
Via delle Industrie 5,
Torre Hammon,
IT-301\,75 Venezia-Marghera,
Italy,~EU
\\
\inst{3}
Swedish Institute of Space Physics,
{\AA}ngstr\"{o}m Laboratory,
P.\,O.~Box~537,
SE-751\,21 Uppsala,
Sweden,~EU
\\
\inst{4}
Galilean School of Higher Education,
University of Padua,
Via VIII Febbraio 1848 n.~2,
IT-351\,22 Padua,
Italy,~EU
}


\abstract{%
We review Heisenberg's uncertainty principle for the orbital angular    
momentum (OAM) of light. By taking into account the largest and        
smallest scales present in nature, such as the the Hubble radius and the 
Planck length, we have found that there exist upper and lower physical  
limits to the determination of the OAM of a photon.                     
}

\begin{document}

\maketitle

\section{Introduction}

Heisenberg's uncertainty principle is one of the most important         
pillars of quantum physics. It describes the intrinsic impossibility    
to measure simultaneously certain physical observables, known as        
conjugate quantum variables. This behaviour is mathematically described 
by the non-commutativity of the mathematical operators associated       
with dynamical variables that do not admit a spectrum of simultaneous   
eigenvalues with an uncertainty on the order of the rationalised Planck 
constant $\hslash$. As is well known, fundamental physical quantities   
such as the coordinate $q$ and its conjugate momentum $p$, and the      
energy $E$ and time $t$ obey, in non-relativistic quantum mechanics,    
the Heisenberg relations ${\Delta{p}\Delta{q}\geq\hslash/2}$ and        
${\Delta{E}\Delta{t}\geq\hslash/2}$. These relations represent the      
basis for the concept of the wave function of a particle.               

Generally, a Hermitian operator $\op{A}$, representing an               
arbitrary physical observable $A$, has the expectation value            
${A_0=\int\cc{\psi}(q)\op{A}\psi(q)\,\d{q}}$, where the probability     
density wave function $\psi(q)$ in most cases is an $L^2$-integrable    
function when the integral converges. The total probability can be      
normalised to unity. In some cases the integral may diverge and the     
probability cannot be normalised in the whole parameter space. However, 
the ratio of the values of the probability at two different points      
of the configuration space can be normalised, thus representing the     
relative probability distribution. The uncertainty in the measurement   
of the quantity $A_0$ is represented by the interval $\Delta{A}$        
defined by $(\Delta A)^2=\int\cc{\psi}(q)(\op{A}-A_0)^2\psi(q)\,\d{q}$. 
If we consider another operator $\op{B}$ associated with the            
observable $B$, the uncertainty principle for the two variables         
$A$ and $B$ is formulated as a general commutator formulation,          
$\bigl[\op{A},\op{B}\bigr]=\im\op{C}\hslash$, so that $\Delta A\,\Delta 
B\geq\frac{\hslash}{2}\Mag{C_0}$. where $C_0$ is the mean value of the  
general commutator $\op{C}$. For conjugate variables generally the mean 
value of the general commutator is $C_0=\pm1$.

\section{Angular momentum of the electromagnetic field}

Classical electromagnetic (EM) radiation can be interpreted in          
terms of an ensemble of photons and the intensity of the radiation      
field is related to the number of photons at a given angular            
frequency $\w=2\pi\nu$. This finds a precise and comprehensive          
description in quantum electrodynamics (QED), the relativistic          
quantum description of the EM field that describes the interaction      
of light with matter \cite{Cohen-Tannoudji&al:Book:1997}. In direct     
correspondence with the second-quantisation formalism of QED, photons   
can alternatively be described in the first quantisation language       
based on the Majorana-Wigner approach using the Riemann-Silberstein     
formalism~\cite{Tamburini&Vicino:PRA:2008}.

EM radiation does not only carry energy $E$ and linear momentum         
$\p$, but also angular momentum $\J$. While $\p$ is connected with      
force action and translational dynamics, $\J$ is connected with         
torque action and rotational dynamics and comprises two distinctively   
different forms \cite{Thide:Book:2011, Thide&al:Incollection:2011}.      
The spin-like form $\S$, known as spin angular momentum (SAM), is       
associated with wave polarisation. The second form $\L$, known as       
orbital angular momentum (OAM), is associated with the phase profile    
of the beam, measured in the direction orthogonal to the propagation    
axis \cite{Beth:PR:1936, Garces-Chavez&al:JOA:2004, Babiker&al:PRL:1994, 
Andersen:PRL:2006}.                                                     

Since a photon does not have a rest reference frame, it is not          
always possible to strictly split the total angular momentum            
$\J$ into two gauge-invariant quantal observables $\S$ and $\L$         
\cite{Berestetskii&al:Book:1989}. The definition of photon spin         
is then derived from general considerations on the quantum              
mechanical properties of the electromagnetic field; in the momentum     
representation, the dependence on the coordinates is replaced by the    
dependence on the momentum $\hslash\k$.\footnote{In a fully covariant   
approach it is found that the photon wave function has a total          
spinorial representation of rank~6, which is equivalent to a spinor of  
rank~2 for each coordinate, namely a vector. For this reason we say     
that the photon has \emph{intrinsic} spin $1$ and the quantum numbers   
associated to helicity are $\lambda=\pm1$, also known as photon spin.}  

The distinguishability of the spin and the orbital angular momentum     
would require that the ``spin'' and ``coordinate'' properties of        
the wave functions be independent of each other, but the photon         
localisability problem makes it impossible to construct, in an          
immediate way, a simultaneous coordinate and momentum representation.   
The vector wave function $\G(\k)$ of the photon must also obey the       
transversality condition, since it is a zero rest mass particle         
\cite{Thide:Book:2011, Thide&al:Incollection:2011}. Because of the       
transversality condition, $\G$ cannot simultaneously specify all the    
values of each of its vectorial components and therefore $\S$ and      
$\J$ cannot be separated. For instance, $\J$ might have a different     
representation in terms of $\S$ and $\L$ when the light beam propagates 
in an inhomogeneous medium.

However, in the case of light beams propagating in vacuum, it is        
possible to separate the two commuting operators $\op{S}_z$ and         
$\op{L}_z $ obtained by projecting the two operators $\Op{S}$ and       
$\Op{L}$ onto the propagation axis of the beam $z$. Letting $\varphi$   
denote the photon wave function, one finds that
\begin{align*}
 &\op{S}_z\varphi = \sigma\hslash\varphi
 &\op{S}^2\varphi = s(s+1)\hslash^2\varphi\,,
\intertext{%
where $s=1$ and $\sigma=\pm1$ ($|\sigma|\leq s$) 
and
}
 &\op{L}_z\varphi = m\hslash\varphi
 &\op{L}^2\varphi = \ell(\ell+1)\hslash^2\varphi\,,
\end{align*}
where $\ell$ and $m$ are integer numbers, and $|m|\leq|\ell|$,          
where $l=0,\pm1,\pm2,\ldots,\pm{}N$. Quantum Electrodynamics            
confirms the picture that each individual photon carries an amount      
of SAM, which is an \emph{intrinsic} property. QED also shows           
that a single photon can additionally carry OAM, which is an            
\emph{extrinsic} property. At the single photon level, the one-photon   
state with OAM can be described either by a Dirac-like equation         
in a superposition of eigenstates of OAM and spin operators, or         
with quantum electrodynamics \cite{Berestetskii&al:Book:1989,           
Calvo&al:PRA:2006}. This property of photons has recently been          
discussed theoretically \cite{Tamburini&Vicino:PRA:2008} and            
confirmed experimentally \cite{Mair&al:N:2001, ONeil&al:PRL:2002,       
Leach&al:PRL:2002, Leach&al:PRL:2004}.

\section{Heisenberg relations for OAM}

The properties of the EM field allow different formulations of the      
uncertainty principle for photons, such as the relationship between     
phase and the photon number or the angular position and orbital         
angular momentum \cite{Beck:93,Vaccaro&Pegg:JMO:1990}. For the          
sake of simplicity and without losing generality, let us consider       
idealised OAM-carrying light beams such as Laguerre-Gaussian (L-G)   
beams that are characterized by helical wavefronts and a                
well-defined $\ell $ value of OAM per photon for any EM frequency       
\cite{Schmitz&al:OE:2006, Thide&al:PRL:2007}. Along the $z$ axis of the  
L-G beam, where the phase is not defined and the field amplitude goes   
to zero, optical vortices (OVs) are found.

The phase is crucial in the OAM states of light, but from the           
quantisation of the electromagnetic field one finds that there is       
no direct formulation of an Hermitian operator for the phase of the     
photon. The construction of a quantum mechanical phase operator for     
the photon exhibits the same difficulties related to the concept of     
angular momentum of an electromagnetic wave as a constant of the        
motion. To construct an Hermitian operator related to the phase one     
approach is to cast two particular operators based on trigonometric     
functions of the phase itself, the ``$\widehat{\mathsf{sin}}$'' and     
``$\widehat{\mathsf{cos}}$'' operators \cite{Mandel&Wolf:Book:1995,     
Kastrup:PRA:2006}. However, they have no immediate physical             
interpretation.                                                         

L-G beams have cylindrical symmetry, and all physical properties        
of such cylindrical systems are periodic functions of an angular        
position. Therefore the angular observables are restricted to           
the range of $2\pi$ and for this reason the angle operator              
$\op{\phi}_\theta$ will have eigenvalues $\phi_\theta$ lying            
in the range $[\theta,\theta+2\pi)$ \cite{Carruthers:RMP:1968,          
Franke-Arnold&al:NJP:2004, Pegg&al:NJP:2005}.\footnote{The subscript    
$\theta$ in the angle operator denotes its dependence on the choice     
of angular range.} The commutator $\op{C}$, associated with the         
formulation of Heisenberg's principle, must be a periodic function of   
the angle $\theta$. For this reason $C_0=1-2\pi P(\theta)$, and the     
ensuing uncertainty relation                                            
\begin{equation}
\label{eq:OAM}
 \Delta\phi_\theta\Delta L_z \geq \frac{\hslash}{2}\Mag{1-2\pi P(\theta)}
\end{equation}
where $P(\theta)$ represents the angular probability density at the     
boundary of the chosen angular range. Following the Wigner-Majorana     
quantisation procedure, photons can be described by a Dirac-like        
equation at the cost of non-localisability of the photon.               

Heisenberg's uncertainty relations were formulated within the realm of  
non-relativistic quantum mechanics. The photon is an ultra-relativistic 
particle so one has to take into account that the speed is limited      
to the speed of light. Let us apply these considerations to the OAM     
states of photons. Already in the 1930's, Landau and Peierls discussed  
Heisenberg's relations when a limit speed must be accounted for so that 
\begin{align}
 (v'-v)\Delta p\Delta t \geq \frac\hslash 2
\end{align}
Because of the existence of a finite limit speed, \viz, the speed       
of light $c$, the absolute value of the (constant) speed difference     
$(v'-v)$ cannot be larger than $c$. In the ultra-relativistic limit,    
when $(v'-v)\sim c$, one obtains a relationship involving momentum      
and time. The coordinate indetermination is then translated into an     
indetermination of the measurement in time,                             
\begin{align}
 \Delta p\Delta t \geq \frac\hslash{2c}\,,
\end{align}
or, in a generic formulation,
\begin{align}
\label{eq:generic}
 \Delta A\Delta t \left(\frac{\Delta B}{\Delta t}\right)
  \geq \frac{\hslash}2 |C_0| \,.
\end{align}

By substituting the quantities in \eqn{eq:OAM} into \eqn{eq:generic},   
assuming $\Delta A=\Delta L_z$ and defining the tangential velocity
around the $z$ axis
\begin{align}
\Omega_\theta= r\ddt{\phi_\theta} \,,
\end{align}
where $\d\phi_\theta/\d{t}$ is the related angular velocity, one        
obtains, for a fixed value of the radius $r$,                           
\begin{align}
\label{eq:7}
 \frac 1r\Delta L_z \Delta t
  \geq \frac{\hslash}{2\Omega_\theta}\Mag{1-2\pi P(\theta)}
  \sim \frac\hslash{2c}
\end{align}
From \eqn{eq:7} one infers that $\Mag{1-2\pi P(\theta)}\sim             
r\Omega_\theta/c$ which means that in a local relativistic limit        
the uncertainty is not determined, since $P(\theta)$ varies in the      
interval $[0,2\pi)$. Multiplying \eqref{eq:7} by $r$, this relation is  
translated into                                                         
\begin{align}
 \Delta L_z\Delta t
  \geq \frac\hslash2\left(\frac {r}{\Omega_\theta}\right)\Mag{1-2\pi P(\theta)}
  \sim \frac\hslash{2}\,\frac rc
 \end{align}
which means that $\Mag{1-2\pi P(\theta)}\sim\Omega_\theta/c$ and in the 
ultra-relativistic limit $r\sim c/\Omega_\theta$. For OAM modes         
\begin{align}
 \Delta L_z= \Delta m\hslash
\end{align}
and Heisenberg's uncertainty relation therefore becomes
\begin{align}
\label{eq:10}
\Delta m\Delta t
 \geq \frac12\left(\frac r{\Omega_\theta}\right)\Mag{1-2\pi P(\theta)}
 \sim\frac12\,\frac rc
\end{align}
This implies that the indetermination of the OAM state combined with
a measurement occurring in the time interval $\Delta t$ must be larger
than half the distance from the phase singularity position in which
one measures the vortex pattern, $r$, divided by the speed of light.    
This is calculated within a phase variation of $2\pi$. The rightmost    
member of \eqn{eq:10} expresses the upper limit obtained in the         
ultra-relativistic case.

\section{Relativistic implications for OAM}

In the usual formulation, an idealised OAM-carrying beam
of light can be represented by the superposition of either           
Laguerre-Gaussian modes or Kummer modes and the dependence of the phase 
of the field does not depend explicitly on the distance to the optical  
singularity. In neither case can the effect of vorticity be measured  
at infinity because Laguerre-Gaussian modes decay exponentially at     
infinity while Kummer beams follow a power-law decay. The entire
wavefront orthogonal to the $z$ axis is twisted in phase and, with
increasing radius $r$, a free test particle would paradoxically move    
around the rotation axis with a superluminal rotation velocity.  This
clearly demonstrates the limitation of the current formalism.

To avoid superluminal velocities one has to consider the r\^ole         
of Special Relativity in the definition of angular momentum while       
formulating OAM states of light. Let us apply a ``\emph{Reductio        
ad Absurdum}'' Gedanken Experiment to prove this conjecture.            
By applying a general coordinate transformation in a Minkowski          
space-time to make $L_z$ disappear locally, we calculate this           
relationship in a co-rotating frame, rotating with angular velocity     
$\d\phi_\theta/\d{t}$ around the propagation axis, $z$, of the light            
beam. In the coordinate set $(t,r,z,\phi)$ the line element in the
Riemannan geometry that describes this particular rotating general        
relativistic flat space-time, is given by the following quadratic form  
\cite{Landau&Lifshitz:Book:1975,Ishihara&al:PRD:1988}                         
\begin{align}
\d{s^2}
 = \left(c^2-\Omega_\theta^2 r^2\right)\d{t^2}
  - 2\Omega_\theta r^2\d\phi\,\d{t}
  - \d{z^2} -r^2\d\phi^2 - \d{r^2}
\end{align}
Even if this metric is locally diagonalisable, because of the           
equivalence principle, it cannot describe the behaviour of a real       
gravitational field at large distances. Otherwise one would have to     
violate the limit of the speed of light for a certain value of the radius 
$r$, violating causality. The mandatory condition on the metric tensor, 
$g_{00}>0$, implies that this particular quadratic form is valid        
only for distances that are in the interval $0<r<c/(\d\phi_\theta/\d{t})$.      
With this new limit imposed by the finiteness of the speed of           
light, one can neglect the rotation of the optical vorticity            
with a local gravitational field only when $r<c/(\d\phi_\theta/\d{t})$,         
paradoxically limiting the spatial extent of the OAM state.

In the ultra-relativistic limit, when the angular velocity               
is $(\d\phi_\theta/\d{t})\simeq c/r$, simple algebra shows that                
$P(\theta)=1/\pi$. By adding the causality condition in a co-rotating   
frame, one finds also an upper limit in the indetermination, namely,    
\begin{align}
 \frac{\hslash}{2\Omega_\theta}\Mag{1-2\pi P(\theta)}
  \leq \frac 1r \Delta L_z\Delta t
  \leq  \frac{\hslash}{2\Omega_\theta}
\end{align}
which implies the following inequalities
\begin{align}
 \frac{r}{2\Omega_\theta}\Mag{1-2\pi P(\theta)}
  \leq  \Delta m\Delta t
  \leq  \frac{r}{2\Omega_\theta}\,.
\end{align}
and the condition
\begin{align}
\label{rquadro}
 \Delta  mc\Delta t= \Delta m\Delta r \sim \frac{r}{2}
\end{align}
This condition reflects the dependence on the radius of the maximum of
intensity and the OAM value in such beams.

Let us assume that a test particle at large distances is rotating       
around the $z$ axis with the angular velocity imparted by photons of    
frequency $\nu=\omega/2\pi$ in an optical vortex of order $m$. In the   
ultra-relativistic case ($v\sim c$) there exists a limit in the OAM 
state indetermination that depends either on the time interval during   
which the local measurement is made                                     
\begin{align}                                                        
 \Delta m \Delta t \sim \frac{r}{c}
\end{align}
or, on the photon wavelength $\lambda_0$,
\begin{align}
 \Delta m \Delta \lambda_0 \sim r
\end{align}

An alternative interpretation can be given in the case of a constant
circular motion with a fixed value of the radius $r$. In this case,
the Heisenberg relation becomes an momentum-angle uncertainty
relation
\begin{align}
 \Delta p \Delta (r \phi_\theta)= r\Delta p\Delta\phi_\theta
  \geq  \frac{\hslash}{2}|C_0|
\end{align}
which is equivalent to the relationship involving the projection onto
the $z$ axis of the OAM operator with the $2 \pi$ periodicity, since $r
\Delta p = \Delta L_z$. We then obtain
\begin{align}
 \Delta m \Delta E \sim \frac c{r}
\end{align}
which means that the maximum indetermination of an optical vortex
cannot be reduced to a point ($r=0$) and, consequently, that any OV must
preserve its central singularity. This clearly reflects the preservation
of the topology of OAM states.

We finally make some remarks about OAM states derived from cosmology.   
The maximum indetermination value in the wavelength estimation of an    
electromagnetic wave, $\Delta\lambda_0|\Sub{max}$, must be smaller      
that the Hubble radius $R=c/H$, where $H$ is the Hubble expansion       
parameter. This defines the size of the universe that has been in       
causal contact with an observer. More precisely, one can assume          
that a photon cannot have a wavelength larger than the radius of the      
last scattering surface, when the universe became transparent to        
radiation, $\Delta\lambda_0|\Sub{max}\sim1.796\times10^{28}$~cm.        
In CGS units one obtains a minimum value of the indetermination of      
the OAM state, and a maximum by selecting the Planck scale, $\Delta     
\lambda_0|\Sub{min}=1.616252(81)\times10^{-33}$~cm, which implies       
a maximum OAM value on the order of $\ell\sim10^{33}$. Hence, the       
finiteness of our universe and the existence of a limit scale such as   
the Planck scale imply the existence of a minimum and maximum value on  
the indetermination of OAM so that
\begin{align}
\label{limit}
 5.5679\times10^{-29} < \Delta\ell < 6.1872\times10^{32}\,.
\end{align}
The indetermination in an OAM state of light will be zero only when the 
Hubble horizon will be infinite, which means an infinite time after  
the Big Bang or with super-horizon modes in an open universe, which     
is, in any case, limited by the last scattering surface. The physical   
meaning of this limit of the indetermination is that, from a classical  
point of view, there can be no sources placed ideally at infinity,      
making the plane wave solution only but an artifact. One may think      
that only spherical modes, according to Huygens's principle, propagate  
in a finite space within a finite time. A different upper limit in      
the indetermination of an OAM state can be derived from superstring     
theory, characterized by a finite string length or from larger scales   
of space-time fuzziness expected from sub-millimetre gravity theories.  
Recent experimental results indicate that the upper limit is closer to   
the Planck scale \cite{tambu2011}.                                      

\section{Conclusions}

From general considerations on Heisenberg's principle for OAM of light  
in a co-rotating frame, we have shown that for the determination of     
OAM states of light, there exist fundamental physical limits dictated   
by the Hubble horizon of the universe, and by the finiteness of Planck  
units below which space and time are not defined. The maximum OAM       
value allowable is of the order $\ell\sim10^{61}$. This is when the     
wavelength is on the order of the observable universe Hubble horizon    
and the twisting step on the order of the Planck scale. A larger        
error might be introduced by the possible presence of submillimetric    
space-time fuzziness expected from quantum gravity theories. In fact,   
one might consider using OAM states for a Gedanken Experiment, either   
to determine the Hubble horizon $H$ or the existence of a scale for     
quantum gravity larger than the Planck scale, by determining the        
boundaries of the indetermination values of OAM states in \eqref{limit}.  
This would represent a direct link from the smallest to the largest     
scales in the universe.


\end{document}